\documentstyle[12pt]{article}
\title{Is there  analogy between  quantized vortex and black hole?}
\author{ G.E. Volovik\\
Low Temperature Laboratory, Helsinki University of Technology\\
Otakaari 3A, 02150 Espoo, Finland\\
and\\L.D. Landau Institute for Theoretical Physics, \\
Kosygin Str. 2, 117940 Moscow, Russia\\}
\begin{document}

\maketitle

\vfill\eject
\begin{abstract}
{An attempt is made to promote an analogy between  quantized vortex in
condensed matter and black hole: both compact objects have fermion zero modes
which induce the finite temperature of these objects. The motion of the
quantized vortex lines in   Fermi  superfluids and superconductors leads to the
spectral flow of fermion zero mode. This results in finite temperature and
entropy of the moving vortex. The tunneling transition rate between the
fermionic levels under the influence of the vortex motion suggests the
effective
temperature of the vortex core $T_{\rm eff}=(2/\pi) p_Fv_L$, where $v_L$ is the
velocity of the vortex with respect to the heat bath reference frame and $p_F$
the Fermi  momentum. This is an analogue of the Unruh temperature of the
accelerating object in the relativistic system. For the vortex ring with the
radius $R$ this leads to the Hawking type temperature $T_{\rm
vortex~ring}=(\hbar v_F /  2\pi R) \ln (R/r_c)$, where $v_F$ is the Fermi
velocity and $r_c$ is the radius of the vortex core. The corresponding
"Hawking" entropy of  the vortex ring of  radius $R$ and  area
$A=\pi R^2$ appears to be ${\cal S}_{\rm vortex ~ring}=(1/6) A p_F^2$. Similar
expression but with different numerical factor is obtained for the instanton
action for the quantum nucleation of the vortex loop from the homogeneous
vacuum, and also for the "Bekenstein" entropy obtained by counting the number
of
the fermionic bound states which appear when the vortex loop is created.
For the superfluid $^3$He-A,  where some components of the order parameter
play the part of the gravitational field, the Fermi momentum $p_F$ corresponds
to the Planck scale for this effective gravity. The effective action for the
gravity field  is obtained after integration over the fermion fields in
correspondence with the Sakharov scenario. The integration over the fermions in
$^3$He-A leads to the renormalization of the vortex entropy and the
"gravitational constant" while their product remains "fundamental". This is the
counterpart of the cancellation of the renormalization corrections to the black
hole entropy and to the gravitational constant discussed by Jacobson.
The statistical-mechanical analysis of the fermions in the vortex core gives
however an essentally less  dynamical  entropy as compared to the
"Hawking-Bekenstein" entropy   and an essentially larger (by the same factor)
temperature of the moving vortex system as compared to the "Hawking-Unruh"
temperature. }
\end{abstract} \vfill \eject

\section{   Introduction}

Quantized vortices in fermi superfluids (and superconductors) provide a
simple example of the macroscopic inhomogeneous system, where the fermionic
excitations move in the collective mean field produced by the motion
of other fermions.  The important property of the mean field potential of the
vortex is that it supports the fermion zero modes -- the gapless or very
low-energy excitations \cite{Caroli}. The fermion zero modes continuously
connect the negative and the positive branches of the energy levels,  which
opens the possibility  for the interaction with the vacuum.  Due to this
contact
some conserved quantities can be transferred from a coherent vacuum motion into
incoherent fermionic degrees of freedom.

In the high energy physics the only realistic objects, which have similar
properties, are the black holes. Other possible objects, such as cosmic
strings, walls and magnetic monopoles, are not topologically stable in the
electroweak vacuum and can occur only in a very high energy scale of the Grand
Unification. Black hole is the macroscopic inhomogeneous system, where the
fermions (neutrons, quarks, etc.) move in the mean (gravitational) field
produced by other particles. The main property of the black hole, which
distinguishes it from  other astrophysical objects such as neutron star, is the
appearance of the fermion zero modes, which open a contact with the "bare"
vacuum. This constitutes the main  analogy with quantized vortices, which we
want to explore here.

It will be shown that the vortex moving  with respect to the heat bath acquires
some effective temperature and entropy (Sec.3), which at first sight (from the
generally looking arguments) have a universal form, similar to the
Hawking-Unruh
temperature and the Hawking-Bekenstein entropy of the black hole
\cite{Hawking}.
This is supported by the consideration of the tunneling rate of the fermions
from the moving vortex (Sec.4.1);  by the $1/R$ and $R^2$ dependence of the
temperature and entropy of the vortex loop of radius $R$ (Sec.4.2); by the
instanton action for the quantum nucleation of the vortex loop from the
homogeneous vacuum (Sec.4.3); and also by the application to the quantized
vortex of the Bekenstein idea on the informational origin of the entropy
\cite{Bekenstein} (Sec.4.4). The information can be lost  when the particle
leaves the positive energy realm. The Bekenstein entropy is calculated by
counting the number of the extra fermionic states which appear when the vortex
loop is created from the homogeneous vacuum state.

However the exact statistical mechanics of the fermions  within the vortex
core (Sec.5) leads to a nonuniversal behavior with  much higher temperature and
much lower entropy of the vortex excitations. The ordering of the fermions due
to  successive occupation of the single-particle states occurs independently on
the information we have about the particles. The same can occur in the black
hole (Sec.6), which can  lead  to the essential reduction of the entropy
compared to the Hawking ansatz.

\section{Spectrum of fermions localized on vortices.}

The quantized vortex is the topologically  stable object of collective motion
in  superfluids and superconductors. The vortex is characterized by
the winding of condensate phase $\Phi(\vec r)$ about the vortex axis and by the
circulating motion of the superfluid component of the liquid around the vortex
with the velocty
$$\vec v_s(\vec r)={\kappa\over 2\pi}\vec\nabla\Phi= N {\kappa\over 2\pi
r}~\hat\phi~~ ,\eqno(2.1)$$
where $\kappa$ is the circulation quantum: $\kappa=\pi\hbar/m_3$ for superfluid
$^3$He and $\kappa=2\pi\hbar/m_4$ for superfluid $^4$He;  $m_3$ and $m_4$ are
masses of the $^3$He and $^4$He atoms; $N$ is the (integer) winding number;
$r$, $\phi$ and $z$ are cylindrical coordinates with the axis $z$ along the
vortex line.

The superflow around the vortex line influences the propagation of
excitations (quasiparticles) near the vortex due to the Doppler effect. In
simplest cases of phonons in superfluid $^4$He and fermions in superfluid
$^3$He-A the classical propagation of these quasiparticles obey the equation of
motion of the scalar wave in the metric \cite{ThreeForces}:
$$ds^2=(c^2-v_s^2(r))(dt +{\kappa\over 2\pi( c^2-v_s^2(r))} d\phi)^2-dr^2-dz^2
-{c^2\over c^2-v_s^2(r)}r^2d\phi^2~~.\eqno(2.2)$$
Here $c$ is the sound velocity for phonon in $^4$He. For the simplest
axisymmetric $^3$He-A vortex the velocity   $c=c_\perp=\Delta/p_F$ enters
which corresponds to the spectrum of fermions in the plane transverse to the
vortex axis. Here $\Delta$ is the gap amplitude  of fermions in bulk $^3$He-A.
The Eq.(2.2) reminds the sonic analogue of black hole discussed in
Ref.\cite{UnruhSonic}. The main difference is however that in the
Ref.\cite{UnruhSonic} the analog of the graviational field is produced by the
normal (nonsuperfluid) motion of the liquid. In our case the gravity is
simulated by the nondissipative superfluid motion with the (superfluid)
velocity
$\vec v_s$. The normal component of the liquid (the system of the
quasiparticles) represents the heat bath and is stationary in the equilibrium,
ie the normal velocity (the velocity of the normal component or the velocity of
heat bath reference frame)
$\vec v_n=0$.

Far from the vortex where $v_s(r)$ is small and can be neglected, this metric
corresponds to that of the so called rotating cosmic string. The spinning
cosmic
string (see the latest references \cite{Rot.String,RotatingString}) is such a
string which has the rotational angular momentum. The metric in Eq.(2.2)
corresponds to the string with the angular momentum  $J= \kappa/ (8\pi G)$ per
unit lehgth and with zero mass.

Approaching the vortex axis one crosses the cylindrical surface of the radius
$$r_c=N\kappa/2\pi c ~~.\eqno(2.3)$$
where the metric has a singularity. This singularity cannot be removed by the
coordinate transformation to a new the reference frame, as is usually made in
the black-hole physics. This is because   the reference  is fixed by the
stationary heat bath. The energy consideration  in the heat bath reference
frame, shows that within the radius $r_c$ (known as the core radius) the
vacuum is unstable and is to be reconstructed.  For Fermi system the
order parameter field in a new stable vacuum is found in
Refs.\cite{KramerPesch,GygiSchluter}. This is made in a self-consistently way,
taking into account the modification of the fermion spectrum in a new vacuum.
Should the vacuum be reconstructed within the horizon of black hole is an open
question.

The quantum mechanical spectrum of single-fermionic excitations on the
background of the vortex contains the states localized near the vortex axis
\cite{Caroli}. The properties of this spectrum  do not depend much on the
detailed structure of the order parameter in the vortex core and are mostly
determined by the topology, ie by the winding number $N$ of the vortex. The
spectrum $\epsilon_{n}(p_z,Q)$ is characterized by the following quantum
numbers: momentum projection $p_z$ on the vortex axis;   the orbital  quantum
number $Q$, integer or half of odd integer, which corresponds to the
generalized
angular momentum conserved in an axisymmetric vortex; $n$ denotes  the radial
quantum number; the spin $s=\pm 1/2$ quantum number is not indicated.

The interlevel distance of the fermionic bound states  $\partial
\epsilon_n/\partial Q=\omega _n$ is usually very small compared to the gap
amplitude:  $\omega_n\sim \Delta^2(T)/E_F\ll \Delta(T)$, where $E_F$ is the
Fermi
energy. Thus for not very small energies, ie in the region $\Delta^2(T)/E_F\ll
\epsilon \ll  \Delta(T)$,  the discrete   $Q$ can be considered as continuous
quantum number. This spectrum has anomalous  (chiral) branches of fermion zero
modes (Fig.1)  whose number $N_{\rm zm}$ is related to the vortex winding
number
$N_{\rm zm}=2N$ according to the index theorem \cite{Q-modes-Index}.  As a
function of (continuous) $Q$, each anomalous branch  crosses zero of energy an
odd number of times and runs through both discrete and contunuous spectrum from
$\epsilon =-\infty $ to $\epsilon =+\infty $. Any other branch either does not
cross zero of energy at all or crosses it  an even number of times. For
low-energy bound states, the spectrum of the chiral branch is linear in $Q$.
For the most symmetric  vortices, for example, this is the branch with $n=0$
\cite{Caroli}
$$\epsilon_0(p_z,Q)=Q\omega_0(p_z)~~ ,\eqno(2.4)$$
this spectrum crosses zero as a function of $Q$ at $Q=0$.

For the continuous vortices in the $^3$He-A the interlevel spacing is inversely
proportional to the core radius in Eq.(2.3) \cite{Kopnin}:
$$\omega_0 \sim {\hbar c \over r_c} ~~ .\eqno(2.5)$$

Due to an odd number of crossings of zero, the spectral flow phenomenon
becomes imortant, which can lead to the creation of the fermions from the
vacuum under an external perturbation. The relevant perturbation is the
motion of the vortex with respect to the heat bath, which constitutes the
normal
component of the liquid.   When the vortex moves with respect to the heat bath
refernce frame, the velocity difference ${\bf v}_n-{\bf v}_L$ induces a flow of
quasiparticles from negative levels to  positive levels of the spectrum
$\epsilon_0(p_z,Q)$. This  results in a  momentum exchange between the moving
vortex and fermions in the heat bath, and thus in the anomalous reactive force
between the vortex and the heat bath \cite{Q-modes-Index,KopninVolovik}. This
force  is a realization of the Callan-Harvey mechanism of the anomaly
cancellation  in the relativistic quantum field theories \cite{Callan-Harvey}.

The spectral flow force depends on the quasiparticle kinetics  determined by
the
parameter  $\omega_0\tau$, where  $\tau$ is the lifetime of fermions.   In the
hydrodynamic limit $\omega_0\tau\ll 1$ the interlevel spacing is smaller than
the level width $1/\tau$ and the spectral flow along the anomalous branch
$\epsilon_0(p_z,Q)$ occurs without any suppression.  When the vortex moves the
angular momentum evolves as $Q\rightarrow Q+({\bf r}(t)\times {\bf
p})\cdot{\hat {\bf z}}=Q+ t(({\bf v}_L-{\bf v}_n)\times {\bf p})\cdot{\hat {\bf
z}}$, with the number
$$\partial_t Q=({\bf v}_L-{\bf v}_n)\cdot ({\bf p}\times{\hat {\bf
z}})~~\eqno(2.6)$$
of levels  crossing  zero energy per unit time. Each level bears the linear
momentum ${\bf p}$, therefore, the total flux of the linear momentum from the
vortex to the heat bath is \cite{Q-modes-Index,KopninVolovik}
$$\partial_t{\bf P}=\sum {\bf p}~(-{\partial f\over \partial Q})\partial_t Q
= -{1\over 2}\sum_{n,Q} {{\partial f(\epsilon_n)}\over {\partial Q}}
 \int_{-p_F}^{p_F} {{dp_z}\over{2\pi}}
\int_0^{2\pi}{{d\phi}\over{2\pi}}~{\bf p}~[ (({\bf v}_L-{\bf v}_n)\times {\bf
p})\cdot{\hat {\bf z}}]=$$
$$  =\pi N {{p_F^3}\over {3\pi^2}}{\hat {\bf z}}\times({\bf v}_L-{\bf v}_n).
{}~~\eqno(2.7)$$

Here we used that only zero modes contribute the sum $-\sum_{n,Q} (\partial
f(\epsilon_n)/\partial
Q)=\sum_n(f(\epsilon_n=-\infty)-f(\epsilon_n=\infty))=2N$  as $Q$ together with
$\epsilon_0(Q) $ run from $-\infty$ to $+\infty $.  Thus the spectral flow
force
between the moving vortex and the heat bath is
$${\bf F}_{\rm sp.flow}=N\kappa\hat {\bf  z}\times C_0({\bf v}_n-{\bf v}_L)
,~\omega_0\tau\ll 1~~.\eqno(2.8)$$
The parameter $C_0$ does not depend on the details of the core structure: it is
expressed in terms of the bulk liquid parameter, the Fermi momentum $p_F$:
$$C_0=mp_F^3/3\pi^2~~, \eqno(2.9)$$
and  coincides with the mass density of the fermi liquid  in the normal
(nonsuperfluid) state.  This independence on the details demonstrates the
topological origin of the spectral flow force.

\section{   Hydrodynamic anomaly due to spectral flow.}

In addition to the spectral flow force, there are   conventional Magnus forces
which act  on the vortex moving with respect to the  superfluid and normal
components:
$$ \kappa\hat {\bf  z}\times  \rho_s(T)({\bf v}_L-{\bf
v}_s(\infty))+\kappa\hat {\bf  z}\times      \rho _n(T)({\bf v}_L-{\bf v}_n)
{}~~.\eqno(3.1)$$
Here ${\bf v}_s(\infty)$ is the constant part of the superfluid velocity
outside the vortex core: the total superfluid velocity around the vortex is
$${\bf v}_s (\vec r)={\bf v}_s(\infty) +   N
{\kappa\over 2\pi r}~\hat\phi~~ .\eqno(3.2)$$
These forces exist even in Bose superfluids where fermions and their
spectral flow are absent. Adding the spectral flow contribution, one obtains
the  balance of forces acting on the moving vortex
$$\kappa\rho_s({\bf v}_s(\infty)-{\bf v}_L)\times\hat {\bf  z}
-\kappa(C_0-\rho_n)({\bf v}_n-{\bf v}_L)\times\hat {\bf  z}  =0~~.\eqno(3.3)$$
Here and further we omit the dissipation, since we are interested in the low
temperature regime,  where the effects of friction, heat conductivity and other
irreversible processes can be neglected.

The hydrodynamic anomaly manifests itself in the so called thermorotation
effect, in which the motion of   vortices induces  the temperature gradient
$\nabla T$. This effect follows from the hydrodynamic equations and
thermodynamic identities \cite{Donnelly}.  The role of the spectral
flow is that it  leads to the temperature gradient even in the
absence of dissipation, ie in the reversible hydrodynamic motion.

For the hydrodynamic description it is relevant to consider an ensemble of
the regularly distributed rectilinear vortices. If $n$ is the
(2-dimensional) density of (singly quantized) vortices then the average
vorticity
of the superfluid velocity is
$$<\vec\nabla\times {\bf v}_s>= \kappa \hat z <\sum_a \delta_2(\vec r-\vec
r_a)> = n\kappa \hat z ~~ , \eqno(3.4)$$
which means that in average the superfluid component  rotates as a solid body
with the angular velocity:
$$\vec\Omega_s={1\over 2} n\kappa \hat z ~~ . \eqno(3.5)$$
Then one has the following
expression for the thermorotation effect in the absence of
dissipation \cite{ThreeForces}:
$$ S\vec\nabla T=   n\kappa  C_0\hat z \times  ({\bf v}_L-{\bf
v}_n)~~ .\eqno(3.6)$$
Thus, if the anomaly parameter $C_0\neq 0$, then even in the absence of
dissipation the entropy and the temperature of the vortices, which  move  with
respect to the heat bath, should be finite.

The distribution of $T(\vec r)$ can be visualized for  the compact object --
the
finite cluster of $N$ singly quantized vortices of radius $R$ (see Fig.2).
According to Eq.(3.5) the radius of the cluster and the angular velocity
$\Omega_s$ of superfluid motion are related by
$$2\pi R^2  \Omega_s=N \kappa  ~~ . \eqno(3.7)$$
In the typical experiments with the vortex cluster the superfluid and normal
component are initially in equilibrium with the rotating container: in this
equilibrium state $\vec\Omega_s=\vec\Omega_n=\vec\Omega_L$. Then the container
is
suddenly stopped and just after stop one has the situation  with
$\vec\Omega_n=0$ and  $\vec\Omega_s\neq 0$, while   the angular velocity
$\vec\Omega_L$ of the vortex lines (with ${\bf v}_L=\vec\Omega_L\times\vec r$)
is  expressed through normal and superfluid velocities according to the force
balance equation (3.3) (see eg \cite{Alles}).

For simplicity let us suppose that the entropy is linear in $T$, ie
$S(T)=\lambda T$: this  situation with finite density of states at zero energy,
which is typical for the Landau Fermi-liquid, is also  valid for the fermion
zero
modes. Then one has $S\vec\nabla T=(1/2)\vec\nabla (ST)$ which gives the
following distribution of $T(\vec r) S(\vec r)$ in the cluster:
$$ S(\vec r) T(\vec r)=   N \Omega_L {\kappa\over \pi}  C_0 {R^2-r^2\over
R^2}~~ . \eqno(3.8)$$
Here it was assumed that $T=0$ outside the cluster. Thus, if the vortex cluster
rotates, ie $\Omega_L\neq 0$, the product of the  temperature and the entropy
density increases towards the center of the cluster reaching the maximal value
$$ S(0) T(0)=N\Omega_L C_0 {\kappa\over \pi}~~   \eqno(3.9)$$
at the center.

This equation does not indicate how $ST$ is distributed between $T$ and $S$.
One should calculate the parameter $\lambda$ in the relation $S(T)=\lambda T$
from the microscopic analysis. This will be considered in the Sec.5, but
before that let us speculate on the temperature and the entropy of the vortex
using some plausible arguments.

\section{   Speculations on vortex entropy and temperature.}

\subsection{Radiation from the moving vortex.}

First let us try to relate the vortex temperature
with the quantum  tunneling of the fermions from the vortex during the vortex
motion \cite{ThreeForces}. The Hamiltonian, which describes the problem at low
$T$ is related only with the low-energy anomalous branch:
$${\cal H}= {\bf Q}\omega_0(p_z)+ \omega_0(p_z)t(\vec v_L\times \vec {\bf
p})\cdot\hat z  ~~.\eqno(4.1.1)$$
Here the second term comes from the change of the angular momentum $Q$ due to
the vortex motion relative to the heat bath, $\vec v_L-\vec v_n$ (we choose
$\vec v_n=0$). The operators of $z$ component of the angular
momentum, $Q$, and transverse linear momentum $\vec p_\perp$ do not commute:
$$[{\bf Q}, \vec {\bf p}_\perp]=i \hat z\times \vec {\bf
p}_\perp~~.\eqno(4.1.2)$$
In terms of the  matrix elements between the states with different $Q$:
$${\cal H}_{QQ'}= Q\delta_{QQ'}\omega_0(p_z)+ \omega_0(p_z)t(\hat
z\times \vec v_L)\cdot <Q\vert \vec {\bf p}\vert Q'>  ~~.\eqno(4.1.3)$$
Here
$$<Q\vert \vec {\bf p}\vert Q'>={1\over 2} p_\perp ((\hat y+i\hat x)
\delta_{Q,Q'+1}+(\hat y-i\hat x) \delta_{Q,Q'-1}) ~~.\eqno(4.1.4)$$

Let us use the semiclassical  approach, which becomes valid, when the vortex
velocity $v_L$ is small compared to $\omega_0 /p_F$. In this case the level
flow
is determined by the exponentially small transition probability between two
neighbouring levels. Let us find this exponent. The Hamiltonian for two level
system, $Q+1$ and $Q$ is
 $${\cal H}=(Q+{1\over 2})\omega_0(p_z) + {1\over 2} \omega_0(p_z)
\left(\matrix{1&v_Ltp_\perp\cr
  v_Ltp_\perp&-1\cr}\right)
\hskip2mm.                                                       \eqno(4.1.5)$$
The square of the energy counted from the position in the middle between the
states  is
$$(E-(Q+{1\over 2}\omega_0(p_z))^2=  {1\over 4} \omega^2_0(p_z)
(1+ (v_Ltp_\perp^2))
\hskip2mm.                                                       \eqno(4.1.6)$$
The trajectory $t=i\tau$ in the imaginary time axis, which connects two states,
gives the following transition probability between the states in the
exponential
approximation:
$$w~\propto ~\exp{-2 {\rm Im}S}~~,~~{\rm Im}S=2\int_0^{\tau_0} d\tau {1\over
2}\omega_0(p_z) \sqrt{1-{\tau^2 \over \tau^2_0}}~~,~~\tau_0={1\over v_L
p_\perp}~~.\eqno(4.1.7)$$
This gives
$$w ~\propto ~\exp{-{\pi\over 2}{\omega_0(p_z) \over v_L
p_\perp}}~~,\eqno(4.1.8)$$
which is equivalent to the thermal distribution  of quasiparticles on the
levels of the anomalous branch of the spectrum with the effective temperature
 $$T_{eff}={2\over\pi} v_L p_\perp ~~.\eqno(4.1.9)$$

It is more pronounced in the 2-dimensional system, where there is no dependence
on $p_z$ and $p_\perp=p_F$ is constant. This temperature equals the
energy of the created quasiparticle  averaged over the azimuthal angle:
  $$T_{eff}= {\overline {\vert \vec v_L\cdot \vec p \vert }}=
{2\over\pi} \int_0^{\pi/2} d\phi v_L p_F \cos\phi ={2\over\pi} v_L
p_F~~.\eqno(4.1.10)$$

If one considers this temperature seriously, then from Eq.(3.9) it follows that
the total entropy of the $N$-vortex cluster is
$${\cal S}\sim \pi R^2 L S(0)\sim N A p_F^2~~,\eqno(4.1.11)$$
where $L$ is the length of
the cluster along $z$ and $A=2\pi R L$ is the surface area of the cluster.
Thus in this reasoning the average entropy of one (singly quantized) vortex in
the cluster, ${\cal S}  \propto p_F^2 A$,  corresponds to the area $A$ swept by
the vortex in its circular motion.

\subsection{Entropy and temperature of the vortex ring.}

For the single  closed  loop of the $N$-quantum vortex with radius $R$ and area
$A=\pi R^2$ the corresponding entropy is similar to the
Hawking entropy  of the black hole with the same radius
$r_g=R$ of the event horizon \cite{ThreeForces}
$${\cal S}_{\rm vortex~ring}= {1\over 6}N Ap_F^2={1\over 4}N
A <p_\perp^2>~~.\eqno(4.2.1)$$
If one takes into account that the vortex ring velocity
$v_L$ and the radius of the ring are related by $v_L=(N\kappa /4\pi R) \ln
(R/r_c)$ one obtains that the temperature $T_{eff}$ of the vortex ring is
$$T_{\rm vortex~ring}={N\kappa p_F\over 2\pi^2 R} \ln {R\over r_c}
{}~~.\eqno(4.2.2)$$
It is inversly proportional to the radius $R$ of the vortex ring in the same
manner as the Hawking temperature of the black hole  is inversly proportional
to
the radius $r_g$ of the event horizon. Similar analogy between the closed
string loop and the black hole was suggested in \cite{Copeland}.

The logarithmic correction in Eq.(4.2.2) disappears in the 2+1 dimensional
case,
where the counterpart of the vortex loop is the pair of oppositely oriented
point vortices. The vortex pair moves with the velocity $v_L=(N\kappa /2\pi
R)$,
where $R$ now is the distance between the vortices in the pair. The
temperature
$T_{eff}$ of the vortex pair is thus
$$T_{\rm vortex~pair}={N\kappa p_F\over \pi^2 R}={N \hbar v_F\over \pi  R}
{}~~,\eqno(4.2.3)$$
where $v_F$ is the Fermi velocity.
This can be compared with the temperature of the black hole   with $R=r_g$
$$T_{\rm black~hole}= { \hbar c\over 4\pi  R}
{}~~.\eqno(4.2.4)$$

The energy of the vortex ring is
$$
E_{\rm vortex~ring}={1\over 2} N^2\kappa^2\rho R \ln {R\over
r_c}~~.\eqno(4.2.5)$$
where the mass density $\rho$ of the superfluid Fermi liquid is very close to
$C_0=mp_F^3/3\pi^2\approx \rho$. As a result
$$E_{\rm vortex~ring}\approx {1\over 2} T_{\rm
vortex~ring}{\cal S}_{\rm vortex~ring}~~.\eqno(4.2.6)$$

In some cases one can relate the Fermi momentum $p_F$ in Eq.(4.2.1) for
the vortex ring entropy to the Planck momentum $p_{Planck}=\sqrt{\hbar
c^3/\gamma}$ in the black hole entropy
$${\cal S}_{\rm BH}= {1\over 4}  A p_{Planck}^2~~.\eqno(4.2.7)$$
This can be done for example in superfluid  $^3$He-A, where some components of
the order parameter play the part of the gravitational field  (see
\cite{Exotic,APhaseGravity}). The effective action for the gravity field
is obtained after integration over the fermion fields. This corresponds
to the Sakharov scenario of the effective gravity \cite{Sakharov}. The
integration over the fermions gives some combination which is equivalent
to the  Einstein-Hilbert term $(1/16 \pi\gamma)\sqrt{-g} R^\mu_\mu$ with the
value of
$$\gamma={\pi \hbar\over 2 p_F^2 c_\perp^2}~~.\eqno(4.2.8)$$
 The factor $c^3$ is absorbed into
the metric tensor: $c_\parallel  c_\perp^2=1/\sqrt{-g}$, where
$c_\parallel=v_F$ and $ c_\perp=\Delta/p_F$  are the velocities of light
propagating along the axis of the axisymmetric vortex and in the transverse
plane correspondingly (see \cite{Exotic,APhaseGravity}).  Thus the Fermi
momentum $p_F$ corresponds to the Planck scale for the effective gravity in
$^3$He-A. This makes the analogy more close.

Note that the equation for the entropy in terms of $\gamma$ does not
depend on the number of species of fermions, since they are absorbed in
$\gamma$. This corresponds to the scenario of Jackobson
\cite{Jackobson} in which the renormalization perturbs both the black hole
entropy and the gravitational constant, while their product remains
fundamental.

\subsection{Instanton action and Hawking entropy of the vortex ring.}

Let us apply to the vortex loop the instanton interpretation of the
black hole entropy. According to \cite{HawkingHorovitzRoss} the instanton
action for the tunneling creation of the pair of the black holes is
proportional to $e^{-{\cal S}_{\rm BH}}$. Let us estimate the tunneling rate of
the nucleation of the vortex loop. The latter is created from the
superfluid vacuum state if the superfluid velocity $v_s$ deviates from the
velocity  $v_n$ of the heat bath. The effective energy of the vortex loop in
the presence of such counterflow $v_s-v_n$ of the superfluid and normal
components of the liquid is (at low temperature)
$$
\tilde E_{\rm vortex~ring}=E_{\rm vortex~ring} - p_{\rm
vortex~ring}(v_s-v_n)=$$
$$={1\over 2} N^2\kappa^2\rho R
\ln {R\over r_c} ~-~\pi N \kappa \rho R^2  (v_s-v_n) ~~.\eqno(4.3.1)$$
where $p_{\rm
vortex~ring}$ is the momentum of the vortex loop which is proportional to its
area. We consider the transition rate between the vacuum state without the
vortex loop ($R=0$) and the vortex state with the same energy
$\tilde E=0$, ie with the the vortex loop of radius
$$
R={N\kappa\over 2\pi (v_s-v_n)}   \ln {R\over
r_c} ~~.\eqno(4.3.2)$$

Such instanton was calculated in \cite{VolovikVortexInstanton} for the  case
when the mass of vortex line was neglected and the quantum nucleation of
the vortex loop was mediated by the irregularity on the surface of
container. Here we follow the arguments of Ref. \cite{Davis} where the quantum
nucleation of vorticity was considered in the homogeneous vacuum and the
existence of the vortex inertial mass $m_{\rm vortex~ring}$ is important. The
result for the semi-classical tunneling is
$e^{-I}$, where
$$I=\int_0^{R} dR^\prime~\sqrt{2 m_{\rm vortex~ring}(R^\prime)\tilde E_{\rm
vortex~ring}(R^\prime) } ~~.\eqno(4.3.3)$$
For the inertial mass of the vortex loop we take the value
$$m_{\rm vortex~ring}={ E_{\rm vortex~ring}\over s^2}~~,\eqno(4.3.4)$$
discussed in \cite{Davis,Duan}, where $s$ is the sound velocity determined by
the compressibility of the liquid with $s=v_F/\sqrt{3}$ for the Fermi gas.
The integration in Eq.(4.3.3) gives again the area law for the entropy of the
vortex loop with the radius $R$ but with different factor:
$$I=\alpha Ap_F^2~~,~~\alpha ={N^2\over 5\pi }\sqrt{32\over 27} \ln {R \over
r_c}~~.\eqno(4.3.5)$$

This nevertheless confirms the suggestion made in  previous subsection that
the vortex entropy can be proportional to the area $A$ of some membrane
terminating on the vortex loop.

\subsection{Bekenstein entropy of the fermions bound to the vortex core.}

Let us consider now the contribution to the entropy from the elementary
excitations in the vicinity of the string loop. We are interesting on the
change of the entropy of the fermion zero modes during creation or annihilation
of the vortex line.

Let us eliminate the vortex loop in the following manner: first one changes the
phase $\Phi$ field in such a way that  everywhere one has $\Phi=0$ except in
the
region within the membrane. The phase $\Phi$ changes by $2\pi N$ when the
membrane is crossed. In this case the membrane represents the $2\pi N$ soliton
(Fig.3). On the second stage one eliminates the soliton together with the
vortex
loop and the homogeneous vacuum state is achieved.

The main property of the soliton is that it gives rise to the fermion zero
modes
whose number is proportional to $N$
\cite{JackiwRebbi,JackiwSchrieffer,SolitonsInPolymers,FermionsIn3HeSoliton}.
The total number of these fermionic bound states
$${1\over 4\pi}N A p_F^2~~,\eqno(4.4.1)$$
which is $N/2$ multiplied by the number of the quantum states of the motion
along the soliton plane:
$2\int dA\int d^2p/ (2\pi)^2=  p_F^2A/2\pi$.

This can be considered as an extra number of states which appeared in the
system when the vortex loop  of the area $A$ is created from the vacuum. Each
zero energy state can be either empty or occupied, which gives the following
Bekenstein \cite{Bekenstein} entropy of fermion zero modes:
$${\cal S} \propto N p_F^2 A ~~.\eqno(4.4.2)$$

We considered the case when the vortex loop (or the vortex-antivortex pair)
annihilates via the soliton wall bounded by the loop.  One can also apply the
Bekenstein arguments using the scenario in which the vortex loop or the vortex
pair annihilates by shrinking. Let us consider a pair of rectilinear vortex
lines with opposite winding numbers $N$ and $-N$. The  anomalous branch, which
enters the vacuum in one vortex,  returns back to the positive energy world in
another vortex. The number of the negative levels is thus $\propto N p_F R$,
where  $R$  is the distance between the vortex and antivortex. Let us suppose
that we loose information on the particle when it enters the negative level
state. This is not so crazy since in the presence of the spectral flow, the
number of  particles occupying the anomalous branch depend on the prehistory.
Since each of these $ p_F R$ states can be either empty or occupied, the
entropy $\propto p_F R \ln 2$. This should be multiplied by the number
$p_F L/\pi$ of longitudinal $p_z$-states, where $L$ is the length of the
vortex.
This again gives the  estimation for the entropy  in terms of the area $A$:
$${\cal S} \propto Np_F^2 RL \ln 2 \propto N p_F^2 A~~.\eqno(4.4.3)$$
This can be considered as the area of the surface swept by
two vortices if they move to each other until complete annihilation: this is
just an another way of the elimination of the vortex pair or of the vortex
loop.

\section{   Microscopic analysis for the vortex temperature.}

However all the arguments above can be applied only in a (semi)classical
macroscopic picture. This means that we   neglected  the energy difference
between the quantum levels of fermions in the core, ie put $\omega_0\rightarrow
0$. Now let us consider an exact statistical-mechanical problem taking into
account the finiteness of the  interlevel distance. In this case one is not
lacking the information on the particles since they successively occupy the
lowest energy levels.  This should essentially reduce the estimated temperature
of the vortex. The same arguments possibly can be applied to the black hole
entropy, which can be essentially  reduced as compared to the
Bekenstein-Hawking
entropy.

Since there are no excitations in the bulk liquid, the temperature gradient
in the vortex cluster should be produced by the vortex core excitations
(fermion
zero modes). Since the vortices rotate with the angular velocity $\Omega_L$,
there is an interaction $Q\Omega_L$ of the rotation velocity with the orbital
momentum $Q$ of the fermion. As a result the distance between the neighbouring
$Q$ levels is
$$\omega_0+\Omega_L~~.\eqno(5.1)$$
The density of the one-dimensional fermionic  states on the anomalous branch
in a single vortex is (if one neglects the dependence of $\omega_0$ on $p_z$ )
$$ {p_F\over \pi}{1\over \omega_0+\Omega_L}~~.\eqno(5.2)$$
The density of states per unit volume is obtained when the Eq.(5.2) is
multiplied by  the  vortex density $n=(2m_3/\pi \hbar)\Omega_s$. This gives for
the density of states per unit volume
$$N(0)= {n p_F\over \pi(\omega_0+\Omega_L)}~~.\eqno(5.3)$$
The energy density of the  fermions is thus
$$E=N(0)\int_0^\infty d\epsilon ~\epsilon~f(\epsilon/T)=n N(0)T^2{\pi^2\over
12}=n T^2 p_F \pi {1\over 12
(\omega_0+\Omega_L)}~~.\eqno(5.4)$$

This gives the parameter $\lambda$ in the relation between the  entropy density
and the temperature
$$S=\lambda T~~,~~\lambda= n   p_F \pi {1\over
12 (\omega_0+\Omega_L)}~~.\eqno(5.5)$$
Thus the rhs of Eq.(3.9) can be distributed between $S$ and $T$ in a different
way depending on ratio $\omega_0/\Omega_L$.

Now one can  apply this to the rotating vortex cluster, discussed in Sec.3.
Equating $S(0) T(0)=\lambda T^2(0)$ to the rhs of Eq.(3.8), ie to $p_F^3n R^2
\Omega_L/3\pi$, one ontains the  temperature in the center of the
cluster  and the entropy of the cluster:
$$T(0)={2\over \pi} R p_F \sqrt{ \Omega_L
(\omega_0+\Omega_L)}~~,~~{\cal S}\sim N p_F^2 A \sqrt{  {\Omega_L\over
 \omega_0+\Omega_L}}~~Ê.\eqno(5.6)$$
In conventional situation one has $\omega_0\gg \Omega_L$ and the entropy is
essentially reduced as compared with the  Bekenstein value. However, in the
(very nonphysical) limit of the small interlevel distance,
$\omega_0\rightarrow 0$, the temperature in the center of the vortex cluster,
expressed in terms of the linear velocity  of vortices on the periphery of the
cluster,
$$T(0)={2\over \pi}  v_L(R)  p_F  ~~.\eqno(5.7)$$
reminds the  vortex temperature in Eq.(4.1.10), while the entropy tends to its
Bekenstein limit.

\section{Statistics of fermions in  black holes and vortices.}

The appartent analogy between   black holes and vortices is in the fermi
distribution of the fermions. In both cases the vacuum is "open", ie the
fermion zero modes appear which connect positive and negative energy
levels. The fermion zero modes are concentrated  within the horizon of the
black
hole and in the core of the vortex and define all the quantum statistics of the
object. The fermions are in the field of the other fermions, which produce the
mean field potential: gravitational field in the black hole and the order
parameter distribution in the vortex core. The fermion zero modes in black hole
can be found in a semiclassical approximation, which is valid for calculations
because the short wave lengths are mostly important.

The radial action, which quantization gives the energy levels $\epsilon(n,L)$
of the fermionic excitations with the bare mass $m$, angular  momentum
$L$ and radial quantum number $n$, is \cite{LandauLifshitz}
$$S_r=\int p_rdr=2\pi n=2\int_0^{r_g} dr~\sqrt{ { \epsilon^2(n,L)\over
c^2(1-{r_g\over r})^2} -
 { {m^2c^2+ {L(L+1)\over r^2}}\over 1-{r_g\over r} } }~~.\eqno(6.1)$$
Here $r_g=2\gamma M/c^2$ is the Schwarzschild radius of black hole with mass
$M$.  At small energy of the excitation $\epsilon\ll mc^2$ the integral is
concentrated in two regions within  the horizon: (1) The vicinity of the
horizon, where the integral is logarithmically divergent at small distances,
gives the contribution
$$S_r^{(1)}=2r_g{\epsilon\over c} \ln   {r_g\over r_0}~~,\eqno(6.2)$$
where $r_0$ is the cut-off parameter (GUT or Planck size). (2) In the region
far from the horizon the term with the energy $\epsilon$ can be neglected as
well as that with $mc^2$ and one has (for $L\gg mcr_g$ )
$$S_r^{(2)} =2L \int_0^{r_g} {dr  \over \sqrt{
 r( r_g -  r)  } }=2\pi L~~.\eqno(6.3)$$
Thus the energy levels are given by
$$ \epsilon(n,L)=\pm \omega_0 (n-L)~~,~~\omega_0={\pi \hbar c\over r_g \ln
{r_g\over r_0} }  = {4\pi^2\over \ln   {r_g\over r_0} } T_{Hawking}
{}~~,\eqno(6.4)$$
where
$$T_{Hawking}=  {\hbar c^3\over 8\pi \gamma M}= {\hbar c \over 4\pi
r_g},\eqno(6.5) $$
is the Hawking temperature.

The Eq.(6.4)  describes the
fermionic zero modes. The interlevel spacing is similar to the Eq.(2.5) for
that of fermions in the vortex core. As distinct from the Eq.(2.4) for the
fermion zero modes in the vortex, which is linear in the generalized angular
momentum $Q$, the zero modes in black holes are linear in two discrete
parameters
$L$ and $n$. The fermion energy becomes exactly zero at  $n=L$, while in the
vortex it is zero at $Q=0$. Also the branches are symmetric, ie their number is
even, as distinct from the excitations on vortices, which have an asymmetric
branch.

The thermodynamics of the black-hole fermions depends on whether  the fermion
number is conserved or not.

(i) If the fermion charge $N_F$ is fixed, then it
determines the mass of the black hole, which equals the total energy of the
fermions. At zero temperature the fermions occupy the lowest energy
levels below the Fermi energy $E_F$ and the total energy is
$$E=Mc^2=\sum_{n,L,L_z; \epsilon<E_F} \epsilon = \sum_{n,L ; \epsilon<E_F}
(2L+1)\epsilon ~~,\eqno(6.6)$$ where $E_F$ is  determined by the
fermionic charge $N_F$:
$$N_F=\sum_{n,L; \epsilon<E_F} (2L+1)  ~~.\eqno(6.7)$$
This gives
$$N_F\sim L^2_{max}{E_F\over\omega_0}  ~~,\eqno(6.8)$$
where the maximal momentum is $L_{max}\sim p_0 r_g=r_g/r_0$, and the Fermi
energy is
$$E_F\sim N_F \omega_0{r_0^2\over r_g^2} \sim N_F \hbar c{r_0^2\over r_g^3}
{}~~.\eqno(6.9)$$
{}From the total energy
$$E=Mc^2 \sim  N_F E_F \sim N_F^2 \hbar c{r_0^2\over r_g^3}
{}~~ \eqno(6.10)$$
one obtains the relation between the fermion number and the mass of the black
hole:
$$M \sim ({\hbar c r_0^2 c^4N_F^2\over \gamma^3})^{1/4}~~ \eqno(6.11)$$

(ii)  Since the vacuum is open, the situation can be the same as in the
superfluid, where within the vortex core there is an exchange of the fermionic
charge between the vacuum and the heat bath of fermions, and thus the fermion
number is not conserved.  In this case the fermion number and the total energy
are nonzero only at nonzero temperature. The total energy of the
system at temperature $T$, which should be equal the mass $M$ of the black
hole,
is given by
$$E=\sum_{n,L,L_z} \epsilon f(\epsilon/T)=\sum_{n,L} (2L+1)~\epsilon
f(\epsilon/T)\sim T^2  L_{max}^2 r_g \ln   {r_g\over r_0} ~~.\eqno(6.12)$$
It was assumed that $T\gg  \pi \hbar c/(r_g \ln  (r_g/ r_0)~)$, ie
$T\gg T_{Hawking}$ otherwise the dependence is exponential. Thus the
estimation of the energy of the black hole with the temperature
$T$ is
$$E=Mc^2\sim T^2 r_g^3 {1\over r_0^2\hbar c}\sim T^2~\gamma^3 M^3
{1\over r_0^2\hbar c^7}~~,\eqno(6.13)$$
ie the temperature of the black hole of mass $M$ is
$$T_{BH}(M)\sim {r_0 c^4\over M} \sqrt{ {\hbar c\over \gamma^3}}\sim {\hbar
c^3\over \gamma M} r_0 p_{Planck}~~,\eqno(6.14)$$
where $p_{Planck}=\sqrt{\hbar c^3/\gamma}$ is the Planck momentum. This
qualitatively agrees with the  Hawking temperature only if the cut-off $r_0$ is
on the Planck  scale. In all realistic  cases  $r_0 p_{Planck}\gg 1$ and
$T_{BH}\gg T_{Hawking}$, which was assumed at derivation. Maybe this means that
in real situation the Hawking limit, where the fundamental temperature is given
by Eq.(6.5), is never reached. Just in the same manner as in Eq.(5.6) the real
fermion temperature of the vortex always exceeds the fundamentally looking
temperature (4.1.10).

The entropy of the black hole also appears to be much less than the
Hawking-Bekenstein entropy. It is zero in the case (i), while in the case (ii)
$${\cal S}_{BH} \sim {\cal S}_{Hawking} {1\over r_0 p_{Planck}
}~~,\eqno(6.15)$$
while ${\cal S}T$, like in vortices,  remains to be invariant and independent
on
the cut-off $r_0$, since it is defined by the black hole mass. The information
on the particles behind the horizon is not lost because of the fermi
statistics,
which allows only one quantum state per each fermion. Each fermion entering the
black hole finally finds a well defined empty state with lowest energy.

The black hole dynamical entropy was discussed in
\cite{FrolovNovikov,Frolov,Barvinsky}. Their result
${\cal S}_{FNB} \sim S_{Hawking} (1/ r_0 p_{Planck})^2$ differs both from the
Eq.(6.15) and the Hawking entropy. It is worthwile to note that their entropy
can be derived from  the same thermal energy $E$ in Eq.(6.13)  as
${\cal S}=2E/T$, if  one uses the Hawking temperature for $T$. However, it
seems that the much higher temperature in Eq.(6.14) should be more relevant.

\section{Conclusion.}

Two systems, black hole and condensed matter vortices, have some similar
features.

(1) In both systems the vacuum is "open", ie they contain the fermion zero
modes, which could lead to the nonconservation of the fermionic charge due to
the spectral flow of the fermions from the vacuum to the heat bath.

(2) The fermion zero modes determine the thermodynamic properties of the
objects: they are responsible for the nonzero temperature of the  vortex, if it
has a finite velocity with respect to the heat bath, and of the black hole with
a finite mass. The black hole moving with respect to the heat bath is to be
considered.

(3) In both systems the entropy is proportional to the area $A$, where $A$ is
the area of horizon in the black hole and in the case of the vortex loop it is
the area of the loop, while the temperature  is $\propto A^{-1/2}$.

(4) Both systems have a limiting case (though possibly not achievable) in which
the temperature and the entropy are given by the fundamental equations. The
temperature is determined by the tunneling of the fermions, while the entropy
corresponds to extra fermionic degrees of freedom which appear when the object
is created from the vacuum state. In the case of the black hole these are the
Hawking-Unruh temperature and the Bekenstein-Hawking entropy.  Similar
expression for the entropy  is obtained for the instanton action for the
quantum
nucleation of the vortex loop and the pair  of black holes from the homogeneous
vacuum.

(5) For the superfluid $^3$He-A,  where some components of the order parameter
plays the part of the gravitational field, one may obtain the dependence of
the vortex entropy on the cut-off parameter. The Fermi momentum $p_F$, which is
the largest momentum in the Fermi liquid theory appears to correspond  to the
Planck scale for the effective gravity in $^3$He-A.  The integration over the
fermions in $^3$He-A leads to the renormalization of the vortex entropy and of
the "gravitational constant" in such a way that their product remains
"fundamental". This is the counterpart of the cancellation of the
renormalization corrections to the black hole entropy and to the gravitational
constant discussed by Jacobson.

\vfill\eject

 \vfill\eject
\begin{figure}
\caption{ Spectrum $\epsilon_n(p_z,Q)$ of the fermions localized in the
condensed
matter vortex in terms of the generalized angular momentum $Q$ and for given
$Q$ in terms of the linear momentum $p_z$ along the vortex axis. The anomalous
branch $\epsilon_0(p_z,Q)$, which crosses as a function of $Q$ the zero energy
level, is the source of the anomaly in the vortex dynamics. }
\end{figure}
\begin{figure}
\caption{ Cluster of the vortex lines. The container and the heat bath rotate
with the angular velocity $\Omega_n$. The superfluid velocity  within the
cluster simulates the solid body rotation of the vacuum (the superfluid
component of the liquid) with the average velocity $<\vec v_s>=\vec
\Omega_s\times \vec r$. If $\Omega_s \neq \Omega_n$, the vortices move with
respect to the heat bath: they rotate as a solid body with velocity  $\Omega_L
\neq \Omega_n$. Due to the spectral flow of fermion zero modes this gives rise
to the finite  temperature of the cluster which  increases towards the center
of
the cluster.}
\end{figure}
\begin{figure}
\caption{ The $2\pi$ soliton - the membrane between the vortex and
antivortex. Outside the soliton the phase $\Phi=0$: the winding of the phase is
concentrated within the soliton. This winding gives rise to the fermion zero
modes within the soliton.}
\end{figure}
\end{document}